\documentclass{Interspeech}
\interspeechcameraready
\title{Rhythm Features for Speaker Identification}

\author[affiliation={1}]{Nick}{Mehlman}
\author[affiliation={2}]{Thomas}{Thebaud}
\author[affiliation={3}]{Dani}{Byrd}
\author[affiliation={1}]{Shri}{Narayanan}

\affiliation{Ming Hsieh Department of Electrical and Computer Engineering}{University of Southern California}{USA}
\affiliation{Department of Electrical and Computer Engineering}{Johns Hopkins University}{Maryland, USA}
\affiliation{Department of Linguistics}{University of Southern California}{USA}
\email{nmehlman@usc.edu}

\usepackage{comment}

\begin{document}

\maketitle

\begin{abstract}
While deep learning models have demonstrated robust performance in speaker recognition tasks, they primarily rely on low-level audio features learned empirically from spectrograms or raw waveforms. However, prior work has indicated that idiosyncratic speaking styles heavily influence the temporal structure of linguistic units in speech signals (rhythm). This makes rhythm a strong yet largely overlooked candidate for a speech identity feature. In this paper, we test this hypothesis by applying deep learning methods to perform text-independent speaker identification from rhythm features. Our findings support the usefulness of rhythmic information for speaker recognition tasks but also suggest that high intra-subject variability in ad-hoc speech can degrade its effectiveness.
\end{abstract}

\section{Introduction}

In addition to its linguistic content, human speech contains rich information about the speaker's identity \cite{hansen_tutorial}. Attributes such as pitch (i.e., $f_0$), spectral energy distribution, and amplitude envelope are highly individualistic and hence can act as useful discriminative features for determining who is speaking (speaker identification, SI) or verifying the identity of a given speaker (speaker verification, SV). However, far less attention has been given to the use of rhythm and other prosodic information, particularly the temporal structure of linguistic units such as phonemes or syllables. This is despite the fact that the rhythmic aspects of speech have been shown to be highly idiosyncratic (e.g., \cite{Igras2014, Pfitzinger2002}) and dependent predominantly on the speaker \cite{Loukina2013}. Additionally, unlike lower-level audio features, rhythm information is more robust to changes in the channel (e.g., noise, filtering) \cite{Mary2008ProsodicFeatures} and even deliberate attempts to obfuscate identity information (e.g., pitch shifting) \cite{tomashenko2024analysis}. While several prior works \cite{vanHeerden2007, bartkova2002prosodic} have attempted to leverage rhythm for SI and SV, most of these methods employ classical learning methods such as hidden Markov models (HMMs). Few attempts have been made to apply more contemporary deep neural networks (DNNs) despite their widespread proliferation and state-of-the-art performance.

In this paper, we explore the utility of rhythm features for DNN-based SI. Our approach uses a transformer model trained on aligned transcripts that are generated automatically from an ASR model. This makes our method text-independent, allowing it to be applied to unlabeled speech signals. We evaluate our method on LibriSpeech \cite{panayotov2015librispeech} and VoxCeleb$1$ \cite{nagrani2017voxceleb}, two popular speech datasets. In addition to the case in which rhythm is used in isolation, we also evaluate its use as an additional feature set to improve the performance of conventional (audio-based) SI models. 

\section{Background}

A variety of prior works have demonstrated that the rhythmic and prosodic aspects of speech convey substantial information about the speaker's identity. For example, \cite{helander2007} demonstrated that human listeners were able to identify familiar speakers based on only a sinusoidal encoding of the prosodic information in speech. Phoneme durations have been shown to be heavily speaker-dependent by both \cite{Igras2014} and \cite{Pfitzinger2002}. Additionally, both \cite{zora68554} and \cite{Loukina2013} observed that these prosodic variations tend to be fairly stable for a given speaker, regardless of the text that is being spoken. Other works, however, such as \cite{Wiget2010HowSA} have found that content can have a substantial influence on certain aspects of the speaker's prosody. 

Based on these observations, a few attempts have been made to leverage rhythmic features to improve the performance of classical speaker recognition methods (e.g., HMMs). For example, \cite{HeerdenNormalization} proposed normalizing phoneme durations to improve the robustness of HMM-based verification systems. However, they found that this produced only relatively small improvements. Meanwhile, the authors in \cite{vanHeerden2007} explicitly used context-dependent phoneme durations for SV. They demonstrated that using these features in conjunction with traditional audio-based ones slightly improved the equal-error rate (EER) compared to audio features alone. Meanwhile, a set of text-independent prosodic features (phone duration, $f_0$, and energy) were individually tested for speaker recognition in \cite{bartkova2002prosodic} and found to carry decent discriminatory power. 

Despite these successes in classical systems, more contemporary SI methods have largely ignored rhythmic information, opting instead to learn empirically derived features directly from the speech signal. Phoneme durations were, however, used to train a transformer-based speaker embedding model in \cite{fujita2021phoneme}. The authors demonstrate that these speaker embeddings produced fairly robust performance in a speaker verification task ($10\%$ EER vs. $2\%$ for x-vectors) and also improved speech synthesis quality. However, their approach used manually aligned annotation, limiting its utility for real-world speech data. Meanwhile, \cite{tomashenko2024analysis} used mean rhythm durations (computed using an HMM-GMM model) as features to attack DNN-based voice anonymization systems. While their approach succeeds in reducing the EER of the anonymized speech, it does not account for potentially context-dependent rhythmic aspects. It also does not directly apply any deep learning techniques to the durational features.

\section{Method}

\begin{figure}[h!]
    \centering
    \includegraphics[width=1.0\linewidth]{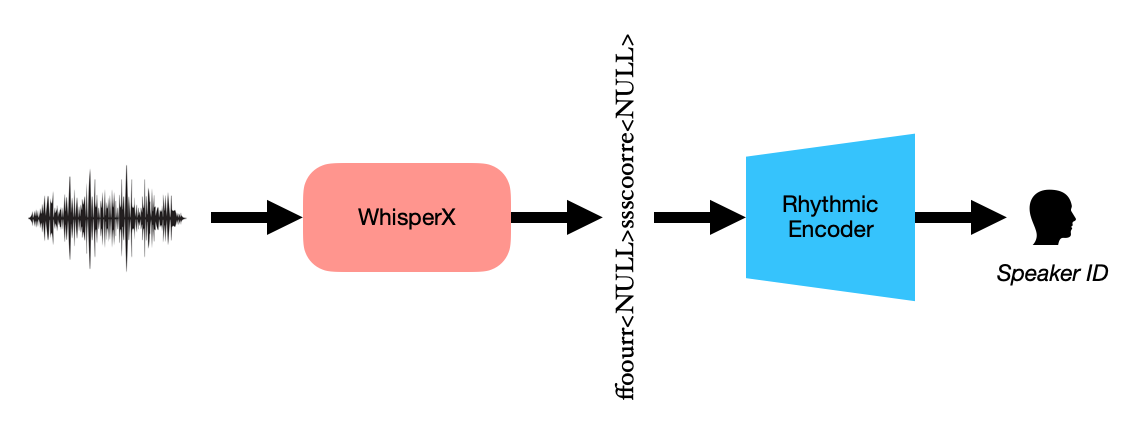}
    \caption{Overviews of the proposed rhythm-based speaker identification framework.}
    \label{fig:overview}
\end{figure}

Figure \ref{fig:overview} summarizes our approach to rhythm-based SI. First, we use the pre-trained WhisperX model \cite{bain2022whisperx} to extract time-aligned transcripts from the audio file. These transcripts are then converted into a frame-aligned character sequence (FACS) with one character per frame and repetition to represent character duration. These FACS are then used to train a transformer encoder model to predict speaker identity. We outline each of these steps in greater detail in the following sections.

\subsection{FACS Extraction}

\begin{table*}[ht]
    \centering
    \renewcommand{\arraystretch}{1}
    \setlength{\tabcolsep}{2pt}
    \begin{tabular}{p{5cm} p{11cm}}
        \toprule
        \textbf{Text} & \textbf{FACS Representation} \\ 
        \midrule
        \textit{`he spoke the last two words'} & hhee**sssppppookee*thee*llaassstt***twwwoo***wwwwworrrddddss \\
        \textit{`in these assemblies we ought to'} & inn***thhhessee***assssssssseemmmbbbllliieess****wwweee*****ouugghht**ttoo \\
        \textit{`she stepped boldly into the room'} & shhee***ssttepppppedd***bbbboollldddddllllyy*iiiiiiiiiiiiiinnnnttooo*thee**rrooooom \\
        \bottomrule
    \end{tabular}
    \caption{FACS representation of different spoken phrases. The null character (which indicated no speech for a given frame) is represented by `*'.}
    \label{tab:facs}
\end{table*}

WhisperX extends the capabilities of the robust Whisper \cite{radford2022whisper} ASR model to generate transcriptions that are time-aligned. To produce our FACS features, we applied WhisperX\footnote{We used the pretrained implementation available at \url{https://github.com/m-bain/whisperX/}} to the speech data to obtain the predicted transcript along with character-level time stamps. These were then converted into a frame-by-frame sequence of characters, with one character per $20$ ms frame. Each character was repeated for the number of frames it was aligned to, thereby capturing the underlying temporal structure of speech. For non-speech frames, we assigned a unique null character to also capture the prosodic information conveyed by pauses and gaps. Note that, unlike prior works, we did not explicitly group characters into phonemes, allowing the rhythm encoder to automatically learn the most useful higher-order linguistic structure. An example of three FACS extracted from the LibriSpeech dataset is shown in Table \ref{tab:facs}. The sequence features for all speech segments were generated offline prior to the training of the rhythm encoder.

\subsection{Rhythm Encoder}

\begin{figure}
    \centering
    \includegraphics[width=0.5\linewidth]{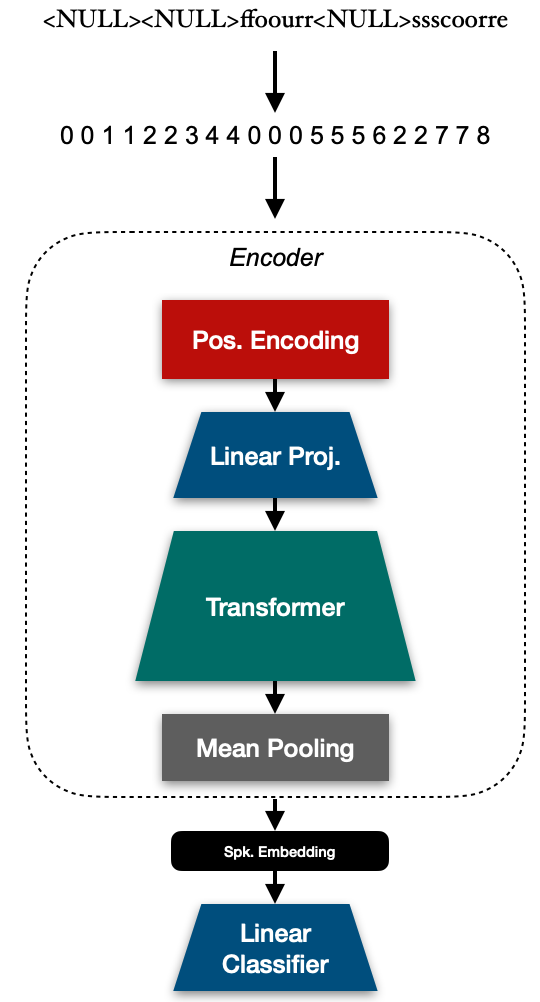}
    \caption{Rhythm encoder architecture.}
    \label{fig:encoder}
\end{figure}

To leverage rhythmic information for speaker recognition, we trained an encoder model, shown in figure \ref{fig:encoder}, to predict speaker identity from the FACS. We elected to use a transformer-based architecture based on these models' success in modeling patterns for sequence-based tasks. In particular, our model consisted of a learnable embedding layer (with one $128$-dimensional embedding per character), followed by positional encoding and $4$ to $6$ transformer layers with $8$-headed attention. In order to limit the model's use of higher-order linguistic features such as words and sentences, we applied attention masking in the transformer layers to restrict the number of adjacent tokens the model could attend to. In practice, we found that using a window size of $\pm 2$ characters produced the best results. The output of the final layer was converted into a one-dimensional vector using mean-pooling across time and then passed through a simple linear head to predict the speaker identity.

\subsection{Rhythm-informed Speaker Identification}

In addition to applying our prosodic model to perform stand-alone speaker identification, we also evaluated its potential to aid the performance of audio-based x-vector models. The x-vectors were extracted offline and then fused with the embeddings from a pre-trained rhythm encoder prior to the classification head. Both the x-vector and rhythmic embeddings were passed through linear projection layers prior to the fusion to address any misalignment between the representation spaces. During training, we jointly updated the linear classifier and the rhythm encoder. 

\section{Experiments}

\subsection{Datasets}

We evaluate our approach on a closed-set SI task using two popular speech datasets.\footnote{For both datasets, we discarded a small number of samples for which the WhisperX alignment failed to converge.} The LibriSpeech dataset \cite{panayotov2015librispeech} consists of clips from English-language audiobook recordings. Each sample is roughly $10-20$ seconds in length and is sampled at $16$ kHz. We used the train-$500$-other split, which contains around $500$ hours of speech data from $1166$ different speakers. The samples were manually subdivided into separate training and testing sets (with split sizes of $90\%$ and $10\%$, respectively), ensuring that the relative representation of each speaker was roughly balanced between the splits. The VoxCeleb$1$ dataset \cite{nagrani2017voxceleb} contains $16$ kHz recordings of speeches of $1251$ celebrities, totaling roughly $350$ hours. We used the pre-existing identification split, which uses roughly $95\%$ of the utterances for training and the remaining $5\%$ for testing. Note that we intentionally selected datasets that represent very different forms of speech (read vs. spontaneous). This allowed us to test the robustness of our approach to different speech contexts. 

\subsection{Rhythm Encoder Training}

The rhythm encoder was trained for $300$ epochs using the Cross-entropy loss function. We used a $4$ layer model on VoxCeleb$1$ and a $6$ layer model on LibriSpeech based on empirical performance during initial training. The batch size was set to $32$, and the initial learning rate was $0.0001$, with a cosine learning rate schedule. Dropout was used to regularize the encoder and mitigate overfitting. We randomly selected $10\%$ of the training split to act as validation data, which was used to determine early stopping based on the model's balanced accuracy. To avoid memory issues, we truncated LibriSpeech FACS at $1024$ tokens and VoxCeleb$1$ FACS at $512$ tokens.

\subsection{X-vector fusion}

We fused the rhythmic embeddings with x-vectors extracted from the pretrained WavLM \cite{chen2022wavlm} and SpeechBrain \cite{speechbrain} speaker recognition models\footnote{These models can be found at \url{https://huggingface.co/microsoft/wavlm-base-sv} and \url{https://huggingface.co/speechbrain/spkrec-xvect-voxceleb} respectively.}. WavLM is a transformer model that was pretrained on LibriSpeech using self-supervised learning and fine-tuned with an x-vector head on VoxCeleb$1$. The SpeechBrain model has a TDNN architecture and was trained on both VoxCeleb$1$ and VoxCeleb$2$.

The joint x-vector and rhythm models were trained for $300$ epochs with the rhythm encoder unfrozen. As a baseline, we also trained linear classifiers on the x-vectors only for $150$ epochs. In both cases, we used an initial learning rate of $0.001$ with cosine scheduling and early stopping, using $10\%$ of the training data for validation.

\section{Results}

\begin{table*}[ht]
    \centering
    \renewcommand{\arraystretch}{1.2}
    \setlength{\tabcolsep}{8pt}
    \begin{tabular}{l|cc|cc|cc}
        \toprule
        & & & \multicolumn{2}{c|}{\textbf{X-vector Only}} & \multicolumn{2}{c}{\textbf{X-vector + Rhythm}} \\ 
        \cmidrule(lr){4-5} \cmidrule(lr){6-7}
        \textbf{Dataset} & \textit{Random} & \textbf{Rhythm Only} & \textbf{WavLM} & \textbf{SpeechBrain} & \textbf{WavLM} & \textbf{SpeechBrain} \\ 
        \midrule
        LibriSpeech & $0.0009$ & $0.3901$ & $0.9775$ & $0.9979$ & $0.9725$ & $0.9953$ \\
        VoxCeleb$1$ & $0.0008$ & $0.0326$ & $0.9782$ & $0.9644$ & $0.9736$ & $0.9513$  \\
        \bottomrule
    \end{tabular}
    \caption{Balanced accuracy performance of $(1)$ rhythm features, $(2)$ x-vectors, and $(3)$ rhythm features $+$ x-vectors for speaker recognition tasks. Results are reported for both the VoxCeleb$1$ and LibriSpeech datasets (on the test split), along with two different x-vector models. The random-chance prediction accuracy is shown in the second column.}
    \label{tab:results}
\end{table*}

We evaluated our models using balanced accuracy since it provides a more calibrated measure of performance than standard accuracy in the case of potential class imbalances within the dataset. Additionally, it is better suited to our closed-set identification task than other commonly used measures, such as equal error rate (EER). The balanced accuracy for a $C$-class classification problem is given by 

$$\frac{1}{C}\sum_{c=1}^C \frac{\mathrm{TP}_c}{\mathrm{TP}_c+\mathrm{FN}_c}$$
Where $\mathrm{TP}_c$ and $\mathrm{FN}_c$ represent respectively the number of true positives and false negatives for class $c$ \cite{grandini2020metrics}.

The results of our experiments are summarized in Table \ref{tab:results}, which reports the test-split balanced accuracy for each of the three scenarios we evaluated. The random-chance accuracy is also listed for comparison purposes. The rhythm-only training performance (column $3$) varies greatly between the two datasets. On LibriSpeech, the model obtains a balanced accuracy of nearly $0.4$, which is substantially higher than the random-chance rate of $0.0009$. However, on VoxCeleb$1$, the rhythm-only model obtains a balanced accuracy of only $0.032$. While still better than random, this is nearly an order of magnitude worse than the LibriSpeech results.

Both the WavLM and SpeechBrain x-vector classifiers (table \ref{tab:results} columns $4$ and $5$ reach near-perfect accuracy (i.e., $>96\%$) on both datasets. The WavLM x-vectors perform roughly $1\%$ better on VoxCeleb$1$ but under-perform the SpeechBrain model by $2\%$ on LibriSpeech. The results for the joint x-vector and rhythm features are shown in the final two columns of Table \ref{tab:results}. The performance of these models is very similar to the models trained on x-vectors alone.

\section{Discussion and Conclusion}
\label{sec:conclusion}

\subsection{Discussion}

Our results are consistent with prior works \cite{Igras2014, Pfitzinger2002} that have suggested rhythm features do convey useful information about a speaker's identity. This is evident in the fact that the rhythm-only models are able to achieve well above random-chance accuracy in predicting speaker identity. However, the large discrepancy in performance between the LibriSpeech and VoxCeleb$1$ datasets indicates that the read speech (LibriSpeech) exhibited more stable prosodic identity markers than the ad-hoc speech (VoxCeleb$1$). Rhythmic identity measures do appear to be robust to speech content as was suggested in \cite{Loukina2013}; for example, the LibriSpeech model is clearly able to extract useful representations from distinct text passages. However, the low accuracy on VoxCeleb$1$ indicates that rhythm far less robust to speech \textit{context}. In particular, the VoxCeleb$1$ dataset is compiled from multiple distinct recordings across a variety of different situations. Factors such as mental state, audience, and subject matter may all impact the specific temporal pattern that a speaker adopts, thus introducing greater intra-individual variability. This influence of non-identity factors on prosodic elements is, for example, supported by the results presented in \cite{bhargava2013improving}, where the authors successfully employed rhythm features for speech emotion recognition.

We also observe that the rhythm features do not appear to add additional discriminatory power over the x-vectors alone. This is evident by the fact that the x-vector only and x-vector plus rhythm models perform nearly identically across the different datasets and x-vector models. This may indicate that the x-vector models have already implicitly learned to encode rhythmic information directly from the raw speech signals. It has been suggested in \cite{raj2019probing} and \cite{Peri2020AnEA} that x-vectors do retain some information about the underlying text and the speaking rate. However, given that the x-vectors alone produce such high accuracy, it might also be the case that the more weakly coupled identity information in the prosodic representation is simply of minimal added utility. It would be helpful to evaluate the marginal utility of the rhythm features in cases in which acoustic features are less reliable, for example, with background noise or over-reduced bandwidth channels. In such cases, the rhythm might provide more meaningful benefits as a parallel stream of identity information.

\subsection{Convergence Speed}

\begin{figure}[h]
    \centering
    \includegraphics[width=1.0\linewidth]{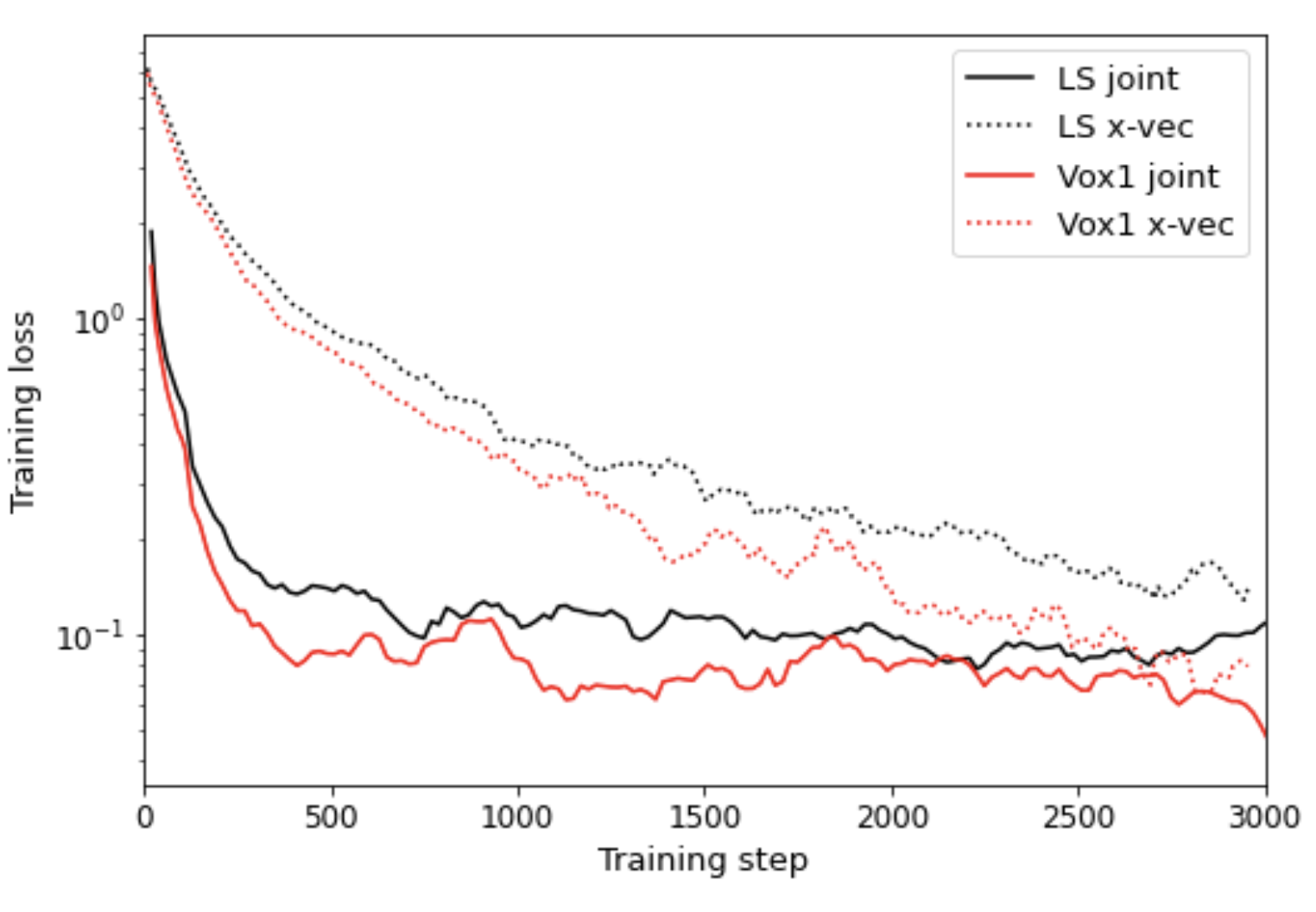}
    \caption{Training loss convergence for WavLM x-vectors only and WavLM x-vectors + rhythm features. Steps are normalized to account for different numbers of training nodes, and loss curves are smoothed using moving-average smoothing with a window of $10$.}
    \label{fig:convergence}
\end{figure}

While the addition of rhythmic information did not produce any benefit in accuracy over x-vectors alone, it did appear to improve convergence rates, especially for the WavLM-derived embeddings. Figure \ref{fig:convergence} shows the training loss over the first $3000$ training steps for the WavLM x-vectors in isolation and for the x-vector/rhythm fusion. The addition of the rhythm embeddings causes the training loss to converge in a smaller number of steps. This trend was also present for the SpeechBrain x-vectors, although less pronounced. 

This finding indicates that the models might leverage the rhythm features more heavily during the initial phases of training when the classifier is less adapted to the x-vector embeddings. In this manner, the rhythm features may still benefit the optimization processes, even if they do not improve the final performance.

\subsection{Limitations}

In this work, we only consider rhythm features in the form of character-level durational representations. We do not include other signal-level prosodic information related to pitch trajectories or energy envelopes. We also focus exclusively on local rhythm features and do not address global temporal information such as speaking rate. Finally, while we apply attention masking to limit the encoder's ability to exploit higher-order linguistic features, it is still possible that some of this information might `leak' into the later layers of the model. Since word choice and sentence structure are also heavily idiosyncratic, this might engender performance benefits not accounted for by rhythmic information alone. 

\subsection{Conclusion}

In this paper, we explored the use of deep-learning-derived rhythm features for SI. Building on prior work that has leveraged prosodic information for classical SR tasks, we applied a pre-trained ASR model to automatically generate frame-aligned transcripts without the need for manual annotation. These transcripts were converted into FACS that captured the temporal dynamics of the speech. Using these FACS, we trained a transformer model to predict speaker identity. Using two well-known speech datasets, we evaluated the efficacy of using rhythm features in isolation and also as a supplement to conventional x-vectors.

Our results indicate that the rhythm features performed well above chance in identification, supporting the hypothesis that speech prosody is substantially individualized. However, there was a large discrepancy in performance between the datasets, indicating that high variability may reduce the utility of rhythm for SI/SV in the case of real-world ad-hoc speech. Additionally, we observed that while the addition of rhythmic information does not produce a performance benefit over x-vectors alone, it does seem to improve the rate of convergence during training.

Future work should consider how rhythmic information might be applicable in low-quality audio channels in which acoustic identity features are less reliable. Additionally, the use of rhythm-only may offer privacy benefits since it removes much of the para-linguistic information, such as gender, age, and health status, that might be discernible from raw audio data.

\bibliographystyle{IEEEtran}
\bibliography{refs}

\end{document}